\documentclass{INTERSPEECH2023}

\interspeechcameraready

\usepackage[utf8]{inputenc} 
\usepackage[T1]{fontenc}    
\usepackage{hyperref}       
\usepackage{url}            
\usepackage{booktabs,arydshln}       
\usepackage{amsfonts}       
\usepackage{nicefrac}       
\usepackage{microtype}      
\usepackage{graphicx}
\usepackage{float}
\usepackage{multirow}
\usepackage{amssymb}
\usepackage{amsmath,amscd}
\usepackage{amsthm}
\usepackage{cancel}     
\usepackage{centernot}  
\usepackage{graphicx}   
\usepackage{soul}       
\usepackage{color}      
\usepackage{enumitem}   
\usepackage{commath}    
\usepackage{subcaption} 
\usepackage{xcolor}
\usepackage{booktabs,arydshln}
\usepackage{CJK}

\title{Prompting the Hidden Talent of Web-Scale Speech Models\\ for Zero-Shot Task Generalization}
\name{Puyuan Peng$^1$, Brian Yan$^2$, Shinji Watanabe$^2$, David Harwath$^1$}
\address{
  $^1$Department of Computer Science, The University of Texas at Austin, USA\\
  $^2$Language Technology Institute, Carnegie Mellon University, USA}
\email{pyp@utexas.edu}

\begin{document}
\begin{CJK*}{UTF8}{gbsn}
\maketitle
 
\begin{abstract}
We investigate the emergent abilities of the recently proposed web-scale speech model \textbf{Whisper}, by adapting it to unseen tasks with prompt engineering.
We selected three tasks: audio-visual speech recognition (AVSR), code-switched speech recognition (CS-ASR), and speech translation (ST) on unseen language pairs. We design task-specific prompts, by either leveraging another large-scale model, or simply manipulating the special tokens in the default prompts. Experiments show that compared to the default prompts, our proposed prompts improve performance by $9\%$ to $45\%$ on the three zero-shot tasks, and even outperform SotA supervised models on some datasets. In addition, our experiments reveal many interesting properties of Whisper, including its robustness to prompts, bias on accents, and the multilingual understanding in its latent space. Code is available 
\href{https://github.com/jasonppy/PromptingWhisper}{here}
\end{abstract}

\vspace{-2mm}
\section{Introduction}
The study of large scale foundation models~\cite{bommasani2021opportunities} has become ubiquitous in many areas of AI, such as large language models for natural language processing~\cite{Brown2020LanguageMA,Wei2021FinetunedLM,wei2022chain}, vision-and-language models for computer vision~\cite{Radford2021LearningTV,zhou2022learning}. 
One of the most intriguing aspects of these large scale pretrained models is their \textbf{emergent ability}~\cite{Wei2022EmergentAO}, usually invoked by \textbf{prompting}~\cite{liu2023pre}, to generalize to unseen data or tasks~\cite{Brown2020LanguageMA,Radford2021LearningTV}. 
In addition to its scientific value, the zero-shot generalization capability of large scale models alleviates the burden of collecting specialized datasets or training special-purpose models for new tasks and domains, resulting in tremendous impact on the application of AI. 

In the field of audio and speech processing, prompt engineering has only recently started to attract attention. 
Gao et al.~\cite{gao2022wavprompt} finetuned a wav2vec2 model ~\cite{Baevski2020wav2vec2A} to produce tokens as prompt for the frozen GPT-2~\cite{radford2019language} to do speech and audio classification tasks. 
Concurrently, Chang et al.~\cite{chang2022exploration} studied gradient-based prompt tuning on a pre-trained speech unit language model~\cite{lakhotia-etal-2021-generative} for speech classification and generation tasks. 
Kim et al.~\cite{Kim2022IntegratedPT} combined learnable prompts and adapters for efficient finetuning of audio models. 
Xue et al.~\cite{Xue2022AWS} is the most similar work to ours. In that paper, the authors trained a Transformer-Transducer model using in-house data on a comparable scale to Whisper, and they ran test time gradient-based adaptation to fine-tune the model for speech translation on unseen language pairs. 
Our work is different from theirs because our adaptation methods are prompt-based and gradient-free, and we study three different zero-shot tasks instead of just one.

\emph{Our work reveals and analyzes the hidden talent and weaknesses of \textbf{Whisper}~\cite{radford2022robust}.} It is the first of its kind that studies \textbf{gradient-free zero-shot task generalization} abilities of web-scale speech models. We show that \textbf{Whisper can be easily adapted to unseen tasks by simply modifying its prompt}. The effectiveness of our proposed prompts are validated on three tasks - \textbf{audio-visual speech recognition (AVSR)}, \textbf{code-switched speech recognition (CS-ASR)}, and \textbf{speech translation (ST) on unseen language pairs}. 

\vspace{-2mm}
\section{The Whisper model}
Here we briefly describe the Whisper model family~\cite{radford2022robust} with an emphasis on the structure of its default prompt.
Whisper is a family of Transformer-based encoder-decoder models~\cite{Vaswani2017AttentionIA} with parameters ranging from 39M (Tiny and Tiny.en) to 1.55B (Large and LargeV2). Whisper models can be categorized into two classes based on languages and tasks: English-only models and multilingual models. The multilingual models are trained on 630k hours of web-scraped speech data for multilingual automatic speech recognition (ASR), En$\rightarrow$X speech translation (ST), language identification (LID), and timestamp prediction. The English models are trained on the English subset of the data (438k hours) for ASR and timestamp prediction.
The encoder of Whisper models takes in log Mel spectrogram, and produces features for the decoder. The decoder consumes encoder features, positional embeddings, and a prompt token sequence. It then produces the transcription of the input speech, or alternatively its translation depending on the prompt. 
The prompt used in the original Whisper paper is the following: \texttt{<|sop|>previous text<|sot|><|language|><|task|><|notimesta mps|>}\footnote{We use \texttt{<|sop|>} to abbreviate \texttt{<|startofprev|>}, and \texttt{<|sot|>} for \texttt{<|startoftranscript|>}. Also \texttt{<|asr|>} for \texttt{<|transcribe|>}, and \texttt{<|st|>} for \texttt{<|translate|>} later.}. Those encapsulated in \texttt{<||>} are special tokens. \texttt{previous text} represents the transcript of the previous utterance, and is optional.
For multilingual models, \texttt{<|language|>} should be replaced by one of the 99 language tokens that Whisper encountered during training. When the input language is unknown at inference, Whisper will first run LID which results in a probability distribution over the 99 languages, and the language with the highest probability is chosen to fill the \texttt{<|language|>} token. 
\texttt{<|task|>} will be replaced by either \texttt{<|asr|>} or \texttt{<|st|>} depending on whether the model should perform ASR or ST. 
We keep \texttt{<|notimestamps|>} in all prompts as our tasks do not need Whisper to produce timestamps\footnote{and therefore we omit this token in the rest of the paper}. 

In all three zero-shot tasks that we consider in this paper, \textbf{we only modify the prompt to the Whisper decoder without modifying the model weights or architecture.} See table~\ref{tab:approach_summary} for a summary of our proposed prompts.
\begin{table*}[ht]
\caption{Summary of our proposed prompts and relative improvement over the default prompts. The differences between our prompt and the default are in \textbf{bold}. In the AVSR task, \texttt{CLIP retrie.} stands for ``CLIP retrieved objects'', and \texttt{<default>} stands for \texttt{<|sot|><|en|><|asr|>}, please find detailed description of our prompt for AVSR in section~\ref{sec:avsr}. For each task only one case is shown in the table, and similar improvements are shown across different datasets and languages in the main text.}\label{tab:approach_summary}
\vspace{-6mm}
\begin{center}
\resizebox{\textwidth}{!}{%
\begin{tabular}{llllc}
\toprule
     Task & Language(s) & Default prompt & Our proposed prompt & Improvement \\
    \midrule
    AVSR & En & \texttt{<|sot|><|en|><|asr|>} & \texttt{\textbf{<|sop|>CLIP retrie.}<default>} & $9\%$ \\
    CS-ASR & Zh+En & \texttt{<|sot|><|zh|>or<|en|><|asr|>} & \texttt{<|sot|>\textbf{<|zh|><|en|>}<|asr|>} & $19\%$ \\
    ST & En$\rightarrow$Ru & \texttt{<|sot|><|ru|><|st|>} & \texttt{<|sot|><|ru|>\textbf{<|asr|>}} & $45\%$ \\
     \bottomrule
\end{tabular}}
\end{center}
\vspace{-7mm}
\end{table*}

\vspace{-2mm}
\section{Audio-visual speech recognition}\label{sec:avsr}
The first task is using an ASR system to produce transcription for a video, where the on-screen visual content is semantically related to the speech audio and can therefore aid in recognition~\cite{sanabria18how2,Gabeur2022Avatar}. This task is related to, but more general than, performing audio-visual speech recognition (AVSR) on speech audio accompanied by a video of the speaker's facial or lip movements~\cite{Chung16}.

\begin{figure}
  \centering
      \includegraphics[width=0.9\columnwidth]{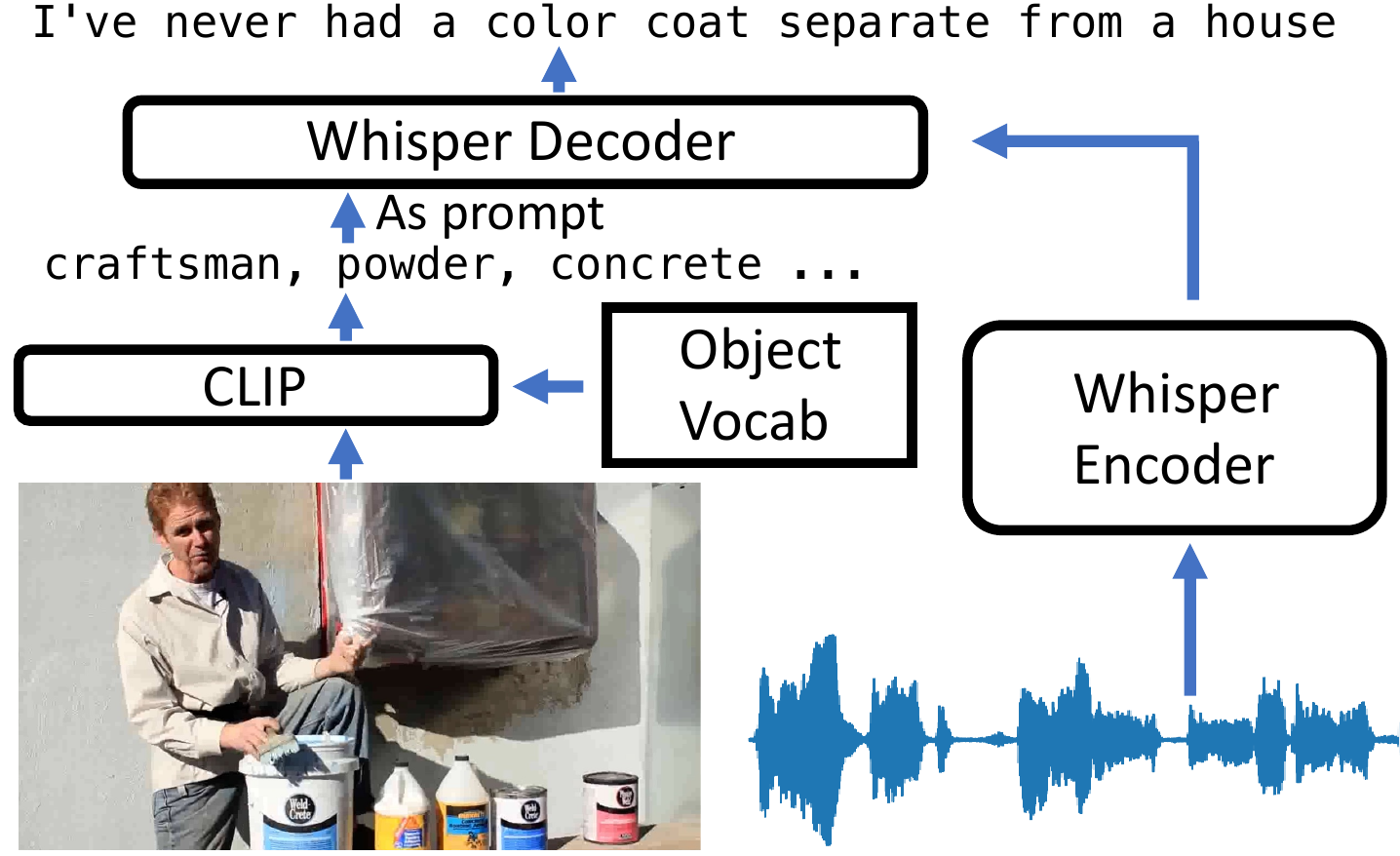} 
      \vspace{-3mm}
      \caption{Framework for visually prompting Whisper. The external object vocab is dataset agnostic.}\label{fig:avsr_visual_prompt}
      \vspace{-5mm}
\end{figure}

\textbf{Approach.} Our approach is shown in figure~\ref{fig:avsr_visual_prompt}.
To provide Whisper with a visually-conditioned prompt, we utilize the popular vision-and-language CLIP~\cite{Radford2021LearningTV} model along with an external vocabulary of common object words to first `convert' the visual stream into a sequence of word tokens. To do so, we take every word/phrase in the external vocabulary, construct a sentence with template ``This is a photo of a \{ \}''. Then we use the CLIP text encoder to pre-compute an embedding vector for each sentence in an offline fashion. At inference time, for each video we sample $3$ equally-spaced RGB image frames, use the CLIP image encoder to embed them, and calculate the similarity between the image embeddings and the pre-computed text embeddings. We select the top $K$ objects whose embeddings have the highest similarity scores with the image embeddings for the prompt.
Next, we concatenate the $K$ selected object names into a comma-separated list of words, and insert this token sequence into the \texttt{previous text} slot of the prompt. 
This method draws inspiration from the idea of Socratic Models~\cite{zeng2023socratic}, where an engineered interface enables large pretrained models to `talk' to one other to solve a complex task.

\textbf{Datasets and implementation details.} Our main dataset for the AVSR task is the recently proposed VisSpeech~\cite{Gabeur2022Avatar}, which is a subset of the instructional video dataset HowTo100M~\cite{miech19howto100m}. 
VisSpeech consists of those videos where an audio-only baseline ASR system performs badly, and whose visual stream and speech audio are semantically related.
Since VisSpeech is proposed as a test set and it only contains 508 examples, we use another instructional video dataset, How2~\cite{sanabria18how2}, for hyperparameter tuning. 
We use a randomly selected 2000 example subset of How2, and add pub noise to the audio to increase the ASR difficulty similar to~\cite{Gabeur2022Avatar}, since the dataset has been shown to be biased towards clean audio~\cite{Gabeur2022Avatar} preventing the visual modality from offering significant benefit to its ASR task. 
For the external object vocabulary, we follow~\cite{zeng2023socratic} and used the label set of Tencent ML-Images~\cite{Wu2019TencentMA}, which contains around 10,000 common objects. The number of object $K$ used in the prompt is tuned for each Whisper model separately on our version of the How2 dataset with three different noise levels (SNR=$5$,$0$,$-5$dB). 

\textbf{Results.}
We found that on our How2 tuning set,  using very large number of objects (as many as $90$ objects) does not hurt performance. Our manual inspection shows that even when using $30$ objects, there are already many irrelevant ones that got mis-retrieved by CLIP. For example, in the example shown in figure~\ref{fig:avsr_visual_prompt}, we found `yogurt', `heavy cream', and `mayonnaise' in the visual prompt. 
In addition, more than $90\%$ of the utterances in our How2 dataset have a ground truth transcription less than $30$ words. 
\emph{This shows that Whisper is very robust to the noise and length of the prompt.} 

\begin{figure*}[h]
  \centering
    \begin{minipage}{0.7\textwidth}
      \includegraphics[width=1.1\columnwidth]{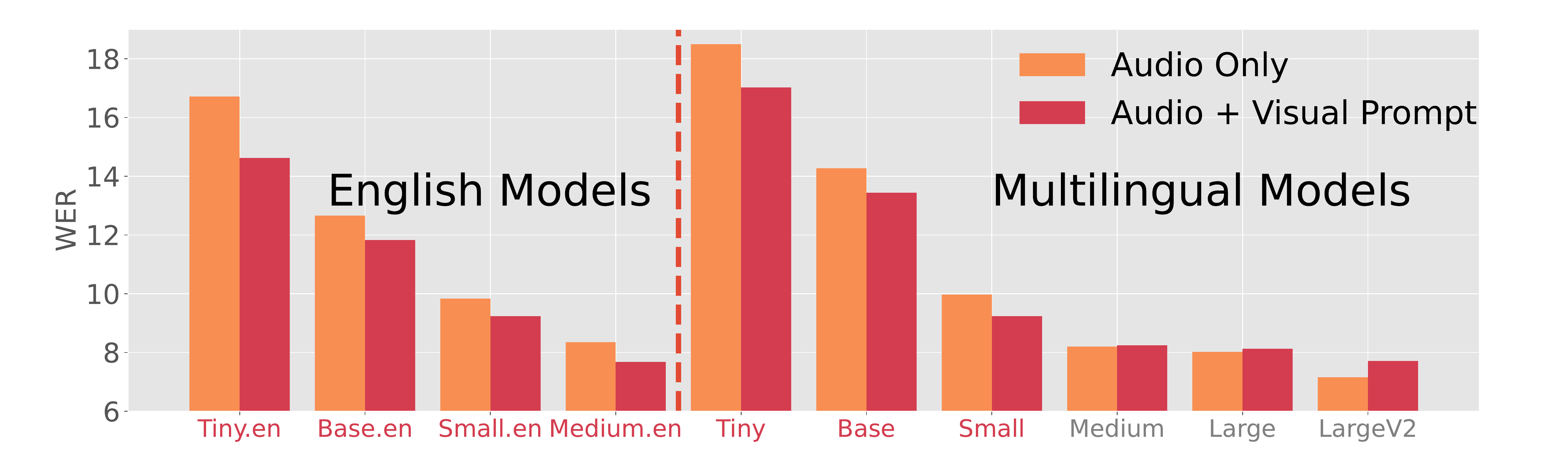} 
      \vspace{-8mm}
      \caption{The effectiveness of visual prompt on VisSpeech across different models.}\label{fig:avsr_visspeech}
  \end{minipage}\hfill
  \begin{minipage}{0.25\textwidth}
\captionof{table}{Comparison of model performance on VisSpeech. With visual prompt, Medium.en outperforms Large.}\label{tab:avsr_tab}
\vspace{-6mm}
    \begin{center}
    \resizebox{\columnwidth}{!}{%
    \begin{tabular}{lcc}
    \toprule
    Model & Modality & WER \\
    \midrule
    SotA~\cite{Gabeur2022Avatar} & A+V & 11.28 \\
    \midrule
    Whisper Medium.en & A & 8.35 \\
    Whisper Medium.en & A+V & \underline{7.60} \\
    \midrule
    Whisper Large & A & 8.02 \\
    Whisper LargeV2 & A & \textbf{7.16} \\
    \bottomrule
    \end{tabular}}
    \end{center}
  \end{minipage}
  \vspace{-6mm}
\end{figure*}

For each model, the top $3$ best performing number of object choices selected from How2 are used for the experiments on VisSpeech\footnote{We use $3$ number of objects choices to reduce the tuning noise introduced by the mismatch between How2 and VisSpeech.}, and the average WER is shown in figure~\ref{fig:avsr_visspeech}. 
We see that the \emph{visually-informed prompt improves the performance for all four English models and three smaller multilingual models, but hurts the performance of the multilingual models Medium, Large, and LargeV2.} 
In table~\ref{tab:avsr_tab}, we compare the previous SotA AVSR results on VisSpeech with the audio-only Whisper performance, and Whisper Medium.en with $50$ objects as the visual prompt. 
We highlight that \emph{visual prompt improve Medium.en by $9\%$, and even outperforms Large.}

\textbf{Remarks.} We propose a prompting approach that that adapts the audio-only Whisper for audio-visual speech recognition. 
Based on figure~\ref{fig:avsr_visspeech}, visual prompting helps most of the models with the exception of three larger multilingual models. However, 
because Large.en and LargeV2.en are not available, it is difficult to draw conclusions on whether it is the model size or multilinguality that hinders the model from benefiting from visual prompting. The fact that visual prompting improves the performance of Medium.en while degrading the performance of Medium suggests that the cause could be multilinguality. If this is the case, multilingual models may benefit from being fine-tuned on monolingual data.

\vspace{-2mm}
\section{Code-switched speech recognition}
Code-switched speech refers to the scenario where more than one language is used in the same utterance. 
With the raise of globalization and democratization of speech recognition technologies, Code-switched ASR (CS-ASR) has become a popular research area~\cite{winata2022decades}.
While we cannot know for certain whether Whisper was trained on code-switched data, it is clear that the model's language and task tokens can not explicitly direct the model to do CS-ASR - each language token only represents one of the 99 training languages, and the task tokens do not convey any information on whether the model should output text in more than one language. 

\textbf{Approach.} To test Whisper's CS-ASR capabilities, we use two Mandarin-English code-switched corpora. See table~\ref{tab:approach_summary} for a quick summary of default and our proposed approach. The default approach (denoted as \texttt{default}) is to let Whisper to first run LID to detect the language between the two\footnote{We set the probabilities of the languages other than the two to be 0.}, and then use the detected language in the prompt. While this default approach work to some degree, it relies heavily on Whisper's LID capabilities, but our results show this can sometimes be inaccurate, especially on accented speech. In addition, this approach doesn't explicitly instruct the model to output text in more than one language for intra-sentential code-switched utterances.
We propose a simple approach called \texttt{concat} that handles the aforementioned issues. The \texttt{concat} approach replaces the single language token in the prompt with two language tokens i.e. \texttt{<|zh|>} and \texttt{<|en|>}, as shown in table~\ref{tab:approach_summary}. 
As will be shown later, despite the simplicity of this approach and the fact that Whisper has never be trained to take two language tokens in the prompt, our approach significantly improves performance.

\textbf{Datasets and implementational details.} We use ASCEND~\cite{lovenia-etal-2022-ascend} and SEAME~\cite{Lyu2010SEAMEAM}, which are both Mandarin-English code-switched datasets. 
Both datasets are spontaneous conversational speech, but ASCEND was recorded from bilingual speakers with different Chinese dialects, while SEAME is recorded from Singaporean and Malaysian speakers. 
We'll show that despite the fact that both datasets contains the same languages,  Whisper performs very differently on them. 
Tuning is done on the validation sets of ASCEND and SEAME. 
For our proposed \texttt{concat} approach, the hyperparameters we tune are: 1. the order of two language tokens in prompt, and 2. the threshold on Whisper's LID confidence score, above which we use the single detected language's token instead of concatenating two language tokens. Whisper Large with \texttt{concat} (\texttt{<|zh|>} first) outperforms all other models and prompts combinations on both datasets, and a threshold of $0.9$ works the best for ASCEND, and $1.0$ i.e. always concatenating two language tokens, works the best for SEAME.

\textbf{Results.} 
Table~\ref{tab:cs_valid} shows the performance of Whisper Large on the validation set of ASCEND and SEAME. In addition to \texttt{default} and our proposed \texttt{concat} prompts, we also show results when we fixed the language token to be \texttt{<|zh|>} or \texttt{<|en|>} for analysis purposes.
\emph{We see that the \texttt{concat} method performs the best on both datasets}, and in particular it provides $19\%$ relative improvement on Total MER (mixed error rate) on SEAME compared to \texttt{default}. 
Secondly, with prompt \texttt{<|zh|>}, Whisper performs much better on pure Mandarin utterances on ASCEND ($16.3$) than SEAME ($26.3$). Similar results are observed for pure English utterances. This indicates that \emph{Whisper's monolingual ASR performance is much worse on SEAME than on ASCEND.} 
Next we note that on SEAME, when we use \texttt{default} instead of \texttt{<|en|>}, En WER increased from $33.8$ to $85.5$, while on ASCEND, WER was $31.8$ in both cases. 
This indicates that i.e. \emph{Whisper's LID performance for detecting English is much worse on SEAME than on ASCEND.}

To understand how do different language prompts steer Whisper's output. We manually examined the error modes, and found a common scenario where \emph{the model outputs monolingual translation for code-switched utterances}. 
This is especially interesting when Whisper does English to Mandarin translation, as the model was only trained to perform X$\rightarrow$En translation. This phenomenon inspired us to quantitatively study En$\rightarrow$X translation capabilities of Whisper in section~\ref{sec:en2x}. 

The test set results for CS-ASR are shown in table~\ref{tab:cs_test}. We see that \emph{with \texttt{concat}, Whisper achieves a new SotA for ASCEND}, while on SEAME there is still an considerable gap between zero-shot Whisper and SotA. 
\vspace{-3mm}
\begin{table}[!ht]
\caption{Performances of Whisper Large on ASCEND and SEAME validation sets. Zh CER shows results on Mandarin utterances, En WER represents results on English utterances. CS MER shows mixed error rate on code-switched utterance. Total MER is the summarizing metric on the entire dataset.}\label{tab:cs_valid}
\vspace{-6mm}
\begin{center}
\resizebox{\columnwidth}{!}{%
    \begin{tabular}{lccccc}
    \toprule
    Dataset & Lang. prompt.  &  Zh CER & En WER & CS MER & Total MER \\
    \midrule
    \multirow{4}{*}{ASCEND} & \texttt{<|zh|>} & \textbf{16.3} & 93.1 & 33.1 & 32.6 \\
    & \texttt{<|en|>} & 90.4 & \textbf{31.5} & 80.1 & 78.9 \\
    & \texttt{default} & 17.0 & \underline{31.8} & \underline{26.6} & \underline{22.1} \\
    & \texttt{concat} & \underline{16.6} & \underline{31.8} & \textbf{25.0} & \textbf{21.3} \\
    \cmidrule(lr){1-6}
    \multirow{4}{*}{SEAME} & \texttt{<|zh|>} & \underline{26.3} & 97.4 & 43.3 & 46.7 \\
    & \texttt{<|en|>} & 99.3 & \textbf{33.8} & 86.9 & 82.2 \\
    & \texttt{default} & 27.1 & 85.5 & \underline{43.2} & \underline{45.3} \\
    & \texttt{concat} & \textbf{25.9} & \underline{44.7} & \textbf{38.4} & \textbf{36.9} \\
    \bottomrule
    \end{tabular}
}
\end{center}
\end{table}
\vspace{-10mm}
\begin{table}[!ht]
\caption{Comparison between \textbf{zero-shot} Whisper Large and \textbf{supervised} SotA models on ASCEND and SEAME test sets}\label{tab:cs_test}
\vspace{-6mm}
\begin{center}
\resizebox{\columnwidth}{!}{%
    \begin{tabular}{lccccc}
    \toprule
    Dataset & Approach  &  Zh CER & En WER & CS MER & Total MER \\
    \midrule
    \multirow{3}{*}{ASCEND} & Sup. SotA~\cite{Nguyen2022OptimizingBN} & - & - & - & 25.0 \\
    & Whisper+\texttt{default} & 19.6 & 30.3 & 23.6 & \underline{22.8} \\
    & Whisper+\texttt{concat} & 16.8 & 30.8 & 22.0 & \textbf{20.9} \\
    \cmidrule(lr){1-6}
    SEAME\textsc{devman}
    & Sup. SotA~\cite{watanabe2018espnet} & - & - & - & \textbf{16.6} \\
     & Whisper+\texttt{default} & 24.7 & 76.3 & 38.2 & 38.2 \\
    & Whisper+\texttt{concat} & 23.6 & 45.8 & 
     33.4 & \underline{32.7} \\
     \cmidrule(lr){1-6}
    SEAME\textsc{devsge}& Sup. SotA~\cite{watanabe2018espnet} & - & - & - & \textbf{23.3} \\
     & Whisper+\texttt{default} & 32.4 & 82.8 & 56.4 & 65.0 \\
    & Whisper+\texttt{concat} & 31.0 & 46.7 & 
     49.6 & \underline{47.6} \\
    \bottomrule
    \end{tabular}
}
\end{center}
\end{table}
\vspace{-5mm}
\begin{figure*}
  \centering
      \includegraphics[width=1\textwidth]{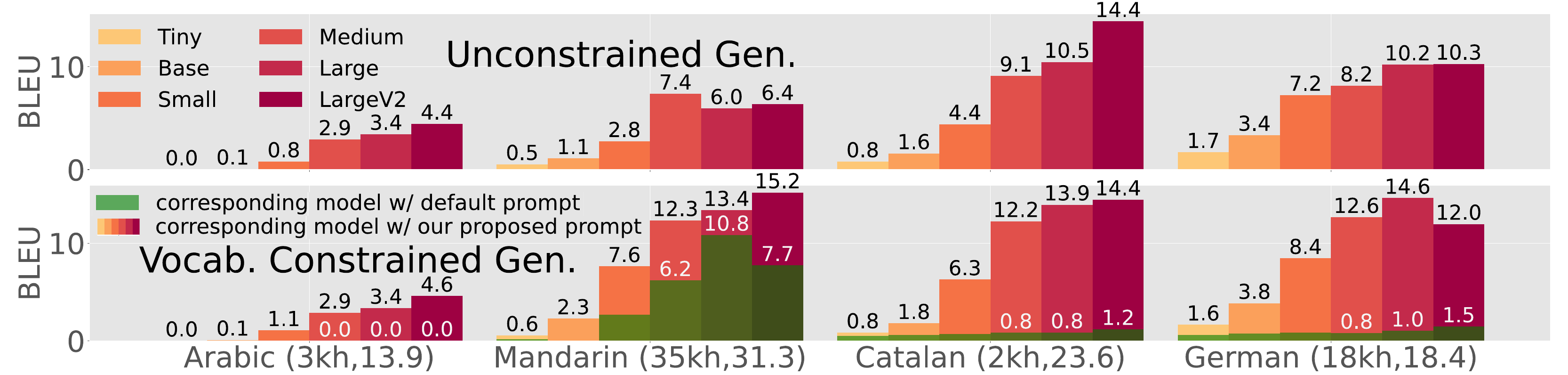} 
      \vspace{-7mm}
      \caption{Whisper Zero-shot En$\rightarrow$X ST results. In the labels of x-axis,  Arabic (3kh,13.9) is interpreted as in the training of Whisper, Arabic ASR data + Arabic$\rightarrow$English ST data amounts to 3k hours, and the supervised ST baseline for English$\rightarrow$Arabic is 13.9. The same interpretation applies to other languages. For the default prompt, we only show numbers for larger models for better visualization. }\label{fig:st_covost2_combined}
      \vspace{-5mm}
\end{figure*}

\textbf{Remarks.}
Recall that ASCEND is Chinese accented, and SEAME is Singaporean and Malaysian accented, and based on our discussion on table~\ref{tab:cs_valid}, we hypothesize that the performance gap on ASCEND and SEAME is because \emph{Whisper's LID and ASR performance vary drastically on different accents}, even though the underlying languages are the same. We leave a more comprehensive investigation of this hypothesis for future work.

\vspace{-2mm}
\section{En to X speech translation}\label{sec:en2x}
In this section, we investigate Whisper's ability to perform En$\rightarrow$X speech translation (ST). Note that Whisper is trained on both multilingual ASR and X$\rightarrow$En speech translation (ST), but never on En$\rightarrow$X ST. Studying Whisper's zero-shot performance on En$\rightarrow$X ST could be a way to measure the speech understanding capabilities that emerge from large scale, multilingual, multitask training. we emphasize that the goal of this section is not to achieve SotA performance, but to study the model's emergent, zero-shot translation ability across different language families, amounts of training data, and model sizes.

\textbf{Approach.} 
The default prompt for ST is to use \texttt{<|st|>} as the task token. However, we found that \texttt{<|st|>} would lead Whisper to only output English no matter what language token is used, unless we constrained the output vocabulary.
To instruct Whisper to do En$\rightarrow$X ST, we propose to use task token \texttt{<|asr|>} instead, and use language token correspond to the language X. See table~\ref{tab:approach_summary} for a example of the default and our proposed approach. 
Counter-intuitive as it might be (using \texttt{<|asr|>} for ST), as we'll show later, our prompt  outperforms the default prompt significantly, and even comes close to supervised approaches for some languages.

\textbf{Datasets and implementation details.} 
We pick Arabic, Mandarin, Catalan and German in CoVoST2~\cite{wang2020covost}, to achieve a resource- and topology-wise diverse evaluation.
To be able to compare with supervised, unsupervised, and other zero-shot ST approaches~\cite{Wang2022SimpleAE}, we also evaluate Whisper on En$\rightarrow$Ru and En$\rightarrow$De from MuST-C V1~\cite{di-gangi-etal-2019-must}, and En$\rightarrow$Fr from Libri-Trans~\cite{kocabiyikoglu-etal-2018-augmenting}.
As for vocabulary constrain, for Arabic, Mandarin, and Russian, we use the unicode range to constrain the vocab to only contain tokens that belong to their scripts; for German, Catalan, and French we constrain the vocab to only contain tokens that are the top $K\%$ most frequent in their training set text. 
$K$ is tuned for CoVoST2 on the development set. For MuST-C and Libri-Trans, we set K to be $40\%$ for German and $50\%$ for French based on CoVoST2 tuning results.

\textbf{Results.} In Figure~\ref{fig:st_covost2_combined}, we show different Whisper models' performance on the four CoVoST2 languages. In general, for our proposed prompt, bigger models perform better across languages, and vocabulary constrained generation outperforms unconstrained generation. As for the default prompt (green bars), we didn't show its performance for unconstrained generation as it only output English text, and for constrained generation, it also performs vert poorly except for Mandarin.
We compare Whisper's performance with other models in table~\ref{tab:st_mustc}\footnote{Whisper Large for En$\rightarrow$De; LargeV2 for En$\rightarrow$Ru and En$\rightarrow$Fr.}. Whisper performs reasonably on all three directions, and especially well on En$\rightarrow$Ru.  
We note that the comparison in this table should only be treated as a reference. This is because even for the unsupervised and zero-shot approaches, they are particularly designed for ST, and they either leverage machine translation systems~\cite{Wang2022SimpleAE,Escolano2020EnablingZM} or multilingual sentence embedding models~\cite{duquenne-etal-2022-modules}. For Whisper, however, we simply adjust its prompt, and the goal is to probe the multilingual understanding of the model.

\textbf{Remarks.} Although Whisper is trained with massive multilingual data, performing En$\rightarrow$X might be harder than one expects. Because for the \texttt{<|st|>} task token, the model is never trained to generate non-English text; for the \texttt{<|asr|>} task token the model is never trained to generate text belonging to a different language than the input speech. 
The fact that Whisper is able to do En$\rightarrow$X ST with a simple modification on its prompt reveals that semantically related words and phrases from different languages might be close in the model's latent space. We also expect that we could fine-tune Whisper to boost the performance of ST on new language pairs.

\vspace{-1mm}
\begin{table}[!ht]
\caption{Comparing zero-shot Whisper with supervised and unsupervised approaches for MuST-C (En$\rightarrow$De and En$\rightarrow$Ru) and Libri-Trans (En$\rightarrow$Fr). Zero-shot Whisper performs reasonably on all three directions. *T-Modules~\cite{duquenne-etal-2022-modules} relies on strong multilingual sentence embedding models that are trained on bitext.}\label{tab:st_mustc}
\vspace{-6mm}
\begin{center}
\resizebox{\columnwidth}{!}{%
    \begin{tabular}{lcccc}
    \toprule
    Category & Approach  &  En$\rightarrow$De & En$\rightarrow$Ru & En$\rightarrow$Fr \\
    \midrule
    \multirow{2}{*}{Supervised} & w2v2+mBART~\cite{Wang2022SimpleAE} &  32.4 &20.0 &23.1 \\
    & E2E Transformer~\cite{wang-etal-2020-fairseq} & 27.2& 15.3 & 11.4 \\
    \cmidrule(lr){1-5}
    \multirow{3}{*}{Unsupervised} & Chung et al.~\cite{chung18} & - & - & 12.2 \\
    & Cascaded~\cite{Wang2022SimpleAE} & 22.0 & 10.0 & 15.4 \\
     & E2E (w2v2+mBART)~\cite{Wang2022SimpleAE} &23.8 &9.8 &15.3 \\
     \cmidrule(lr){1-5}
     \multirow{4}{*}{Zero-shot} & Escolano et al.~\cite{Escolano2020EnablingZM} & 6.8 & - & 10.9 \\
     & T-Modules*~\cite{duquenne-etal-2022-modules} & 23.8 & - & 32.7 \\
     & Whisper w/ default prompt & 0.4& 8.8& 0.8\\
     & Whisper w/ our prompt & 18.1 & 12.8 & 13.1 \\
    \bottomrule
    \end{tabular}
}
\end{center}
\end{table}
\vspace{-6mm}
\section{Conclusion}
We investigate the emergent abilities of Whisper through the lens of prompt-based zero-shot task generalization. 
Our proposed prompts significantly outperform the default prompts in all three tasks that we studied. In addition, we found interesting properties of Whisper - 
in AVSR, we found that the model is very robust to the length and noisiness of the visual prompt, and the effectiveness of the visual prompt between English models and multilingual models are quite different; 
in CS-ASR, we identified potential performance gaps between different accents; 
in ST, we found the surprising results that the \texttt{<|asr|>} task token can be used to instruct the model to do translation and outperforms \texttt{<|st|>}. 
Many of the above properties are worth further investigating, and can potentially lead to models that are more robust, more generalizable, and have less unwanted bias.
\bibliographystyle{IEEEtran}
\bibliography{mybib}
\appendix
\section{Appendix}
\subsection{Examples of CS-ASR transcription}
\begin{table*}[t]
\caption{Examples of how \texttt{concat} improve transcription over \texttt{default} on SEAME with Whisper Large. We use ... when transcriptions are the same for all three cases.}\label{tab:cs_example}
\vspace{-6mm}
\begin{center}
\begin{tabular}{lll}
\toprule
     Ground Truth & transcription w/ \texttt{default} & transcription w/ \texttt{concat} \\
    \midrule
    也 不 需 要 做 research&也 不 需 要 做 研 究&也 不 需 要 做 research\\
    这 真 的 是 一 个 very tough question ... & 这 真 的 是 一 个 很 困 难 的 问 题...& 这 真 的 是 一 个 very tough question...  \\
    每 次 ... 还 是 choir practice & 每 次 ... 还 是 quiet practice&每 次 ... 还 是 choir practice \\
    then did you realise the performances...&那 你 有 没 有 意 识 到 表 演&then do you realise the performances... \\
    ...你 真 的 是 要 睡 觉 了 是 吗 & ...你 真 的 是 要 sweet 小 的 是 吗 & ...你 真 的 是 要 睡 觉 了 是 吗 \\
     \bottomrule
\end{tabular}
\end{center}
\end{table*}

\end{CJK*}
\end{document}